\documentclass[preprint,superscriptaddress,twocolumn,lengthcheck,tightenlines,showpacs,a4paper,10pt,prl]{revtex4}
\usepackage{amssymb}
\usepackage{amsfonts}

\input{tcilatex}
\begin{document}

\preprint{APS/123-QED}
\title{ Spin Liquid State in the Disordered Triangular Lattice Sc$_{2}$Ga$%
_{2}$CuO$_{7}$ Revealed by NMR}
\author{P. Khuntia$^{\ast ,\dagger }$}
\affiliation{Ames Laboratory, US Department of Energy, Iowa State University, Ames, Iowa
50011, USA}
\author{R. Kumar}
\affiliation{Department of Physics, Indian Institute of Technology Bombay Powai
Mumbai-400076, India}
\author{A. V. Mahajan}
\affiliation{Department of Physics, Indian Institute of Technology Bombay Powai
Mumbai-400076, India}
\author{M. Baenitz}
\affiliation{Max Planck Institute for Chemical Physics of Solids, 01187 Dresden, Germany}
\author{Y. Furukawa}
\affiliation{Ames Laboratory, US Department of Energy, Iowa State University, Ames, Iowa
50011, USA}
\affiliation{Department of Physics and Astronomy, Iowa State University, Ames, Iowa
50011, USA}

\begin{abstract}
We present microscopic magnetic properties of a two dimensional triangular
lattice Sc$_{2}$Ga$_{2}$CuO$_{7}$, consisting of single and double
triangular Cu planes. An antiferromagnetic (AFM) exchange interaction $J$/%
\textit{k}$_{\mathrm{B}}$ $\approx $ 35 K between Cu$^{2+}$ ($S$ = 1/2)
spins in the triangular bi-plane is obtained from the analysis of intrinsic
magnetic susceptibility data. The intrinsic magnetic susceptibility,
extracted from $^{71}$Ga NMR shift data, displays the presence of \ AFM
short range spin correlations and remains finite down to 50 mK suggesting a
non-singlet ground state. The nuclear spin-lattice relaxation rate (1/$T_{1}$%
) reveals a slowing down of Cu$^{2+}$ spin fluctuations with decreasing $T$
down to 100 mK. Magnetic specific heat (\textit{C}$_{m}$) and 1/$T_{1}$
exhibit a power law behavior at low temperatures implying gapless nature of
the spin excitation spectrum. Absence of long range magnetic ordering down
to $\sim J$/700, nonzero spin susceptibility at low $T$, and power law
behavior of \textit{C}$_{m}$ and 1/$T_{1}$ suggest a gapless quantum spin
liquid (QSL) state. Our results demonstrate that persistent spin dynamics\
induced by frustration maintain a quantum-disordered state at $T$ $%
\rightarrow $ 0\ in this triangular lattice antiferromagnet. This suggests
that the low energy modes are dominated by spinon excitations in the QSL
state due to randomness engendered by disorder and frustration.
\end{abstract}

\volumeyear{year}
\volumenumber{number}
\issuenumber{number}
\eid{identifier}
\date[Date text]{: updated version on: }
\received[Received text]{date}
\revised[Revised text]{date}
\accepted[Accepted text]{date}
\published[Published text]{date}
\startpage{1}
\endpage{2}
\maketitle

Collective excitations, frustration, and quantum fluctuations are key
ingredients in driving novel ground state properties of correlated electron
systems. Geometrically frustrated magnets harbor exotic physical phenomena
such as spin glass, quantum spin liquid (QSL), spin ice, and
superconductivity \cite{LB, PA,SS,CL,HTD}. The incompatibility of magnetic
exchange interactions in achieving minimum energy yields degenerate ground
states and the associated strong quantum fluctuations prevent the spin
system from undergoing a symmetry breaking phase transition \cite%
{LB,SS,CL,HTD,STB,RMR,JSG,CC}. The experimental realization of novel states
such as QSL in real materials is an exciting prospect in answering some of
the key issues in condensed matter and set an enduring theme following
Anderson's resonance valence bond theory \cite{PWA,PWS}. The most prominent
QSL candidates reported so far are $S$\textit{\ }= 1/2 kagom\'{e} lattices
ZnCu$_{3}$(OH)$_{6}$Cl$_{2}$, Cu$_{3}$Zn(OH)$_{6}$Cl$_{2}$, [NH$_{4}$]$_{2}$%
[C$_{7}$H$_{14}$N][V$_{7}$O$_{6}$F$_{18}$], $S$\textit{\ }=1/2 hyperkagom%
\'{e} \ PbCuTe$_{2}$O$_{6}$, Na$_{3}$Ir$_{4}$O$_{8}$, and organic $S$ = 1/2
triangular lattice, EtMe$_{3}$Sb[Pd(dmit)$_{2}$]$_{2}$, $\mathit{\kappa }$%
-(BEDT-TTF)$_{2}$Cu$_{2}$(CN)$_{3}$, and $\kappa $-(ET)$_{2}$Cu[N(CN)$_{2}$%
]Cl. The spin excitation spectra in the QSL state can be gapped or gapless
with exicitng magnetic properties \cite%
{CL,TH,NPS,JS,PM1,AO,TI,TI2,BF,LC,SED,PMRev,YO,AP,YSZ,FLP,SY,RK,SYQ,SY1,AEA,OT,DW,XGW,PKH}%
. The frustrated quantum magnets are proposed to host emergent fractional
excitations in the gapless QSL state, which is reflected as power law
behavior in bulk and microscopic observables \cite%
{LB,ZY,Th1,PAL1,PAL2,ST,ST1,EKH}. Recently, the observation of intriguing
magnetic properties in Ba$_{3}$TSb$_{2}$O$_{9}$ (T=Cu, Co, Ni) and 5\textit{d%
} iridates has rekindled enormous research activities in quantum materials
in the context of emergent quantum states \cite%
{LB,PA,YO,HSZ,SN1,PM3,S1,PK1,YSL,NP1}. Among the frustrated magnets, the
edge-shared triangular lattice AFM with $S$ = 1/2 offers the simplest
archetype for QSL and to test theoretical models in other relatively complex
lattices \cite{NPS,FLP,RK,SYQ}. Furtheremore, the role of intersite
distribution or disorder in stabilizing a QSL state in frustrated quantum
magnets has recently been suggested \cite{BF,PK1,TF,FM}.

In view of the vastly evolving field of frustrated magnetism, significant
attention has recently been paid to the growth and design of new quantum
magnets which could epitomize as model materials for hosting exotic
excitations pertinent to novel states and to test theoretical conjectures 
\cite{LB,CL,HTD,CC}. In the quest for novel states in frustrated magnets
with low spin where inherent quantum effects lead to emergent phenomena, we
synthesize and investigate an inorganic $S$ = 1/2 antiferromagnet Sc$_{2}$Ga$%
_{2}$CuO$_{7}$ (henceforth SGCO). Recent detailed synchrotron x-ray and
neutron diffraction measurements revealed that the magnetic lattice
comprises of triangular bi-planes of correlated Cu$^{2+}$ spins dominated by
50 \% Ga$^{3+}$ ions due to unavoidable intersite inversion and the single
triangular plane is mainly occupied by non-magnetic Ga$^{3+}$ ions and 15 \%
Cu$^{2+}$ in the single triangular plane give rise to a paramagnetic
behavior. The bulk magnetic susceptibility at low temperature is dominated
by defect contributions and specific heat displays no signature of long
range ordering down to 0.35 K, which invokes microscopic investigations \cite%
{SCGO-2}. Absence of significant anisotropy and no appreciable spin-orbit
coupling suggest that SGCO might be a promising quantum magnet to address
low lying excitations intrinsic to the triangular lattice.

The microscopic details pertaining to the magnetic properties inherent to
the magnetic lattice at very low temperature is a very crucial step forward
for establishing the ground state convincingly and in exploring the nature
of low lying excitations. Herein, we report the first nuclear magnetic
resonance (NMR) studies on a new $S$ = 1/2 inorganic triangular lattice
SGCO. NMR being a powerful local probe sheds light on the intrinsic spin
susceptibility and the dynamic spin susceptibility via spectra and
spin-lattice relaxation rate (1/$T_{1}$) measurements, respectively, from a
microscopic point of view. The intrinsic spin susceptibility suggests the
presence of AFM\ spin correlations with $J$/\textit{k}$_{B}$ $\approx $ 35 K
\ between Cu$^{2+}$ spins in the triangular bi-planes and non-singlet state
without signature of long range magnetic ordering (LRO) down to 50 mK. The 1/%
$T_{1}$ data suggest a slowing down of Cu$^{2+}$ spin fluctuations with
decreasing temperature down to 100 mK and power law behavior of magnetic
specific heat (\textit{C}$_{m}$) and 1/\textit{T}$_{1}$ imply gapless spin
excitations. Our comprehensive results establish a gapless quantum spin
liquid state in SGCO.

Polycrystalline sample of Sc$_{2}$Ga$_{2}$CuO$_{7}$ was prepared by a method
described elsewhere \cite{SCGO-2}. SGCO crystallizes in a hexagonal
structure with a space group $P6_{3}/mmc$ and lattice constants $a=b=$
3.30479(4) \AA\ and $c$ = 28.1298(4)\AA\ . The magnetic structure comprises
of alternating single and double triangular planes. The interaction between
the Cu$^{2+}$ spins is confined to the 2D triangular bi-plane only, with
negligible interlayer interactions \cite{SCGO-2}.

Shown in Fig. 1(a) is the temperature dependence of bulk magnetic
susceptibility $\chi _{\mathrm{obs}}$, which is found to be strongly
enhanced at low temperatures without exhibiting any signature of long range
magnetic ordering (LRO) down to 1.8 K. We did not observe ZFC and FC
splitting in $\chi _{\mathrm{obs}}$ and no hysteresis was found in
magnetization \cite{SCGO-2}. The green dotted line in Fig. 1(a) shows the
magnetic susceptibility $\chi _{\mathrm{sub}}$ after subtracting from $\chi
_{\mathrm{obs}}$\ a contribution due to the presence of 15\% Cu spins on the
triangular plane assuming a simple Curie behavior of $S$ = 1/2 for the Cu
spins. The Curie-Weiss (CW) fit of $\chi _{\mathrm{sub}}$ at high
temperatures above 100 K yields a CW temperature $\theta _{\mathrm{CW}}=-$44
K, an effective magnetic moment ($\mu _{eff}$) of 1.83 $\mu _{\mathrm{B,}}$
and $g$ $\approx $\ 2. The negative value of $\theta _{\mathrm{CW}}$
indicates the presence of AFM interaction between Cu$^{2+}$ spins on the
triangular bi-plane. The \textit{T}-dependence of \ magnetic specific heat
(as shown in Fig. 1(b)) in different magnetic fields don't display any sign
of LRO. The magnetic specific heat (\textit{C}$_{m}$) exhibits a power law ($%
\sim $\textit{T}$^{1.9}$) behavior indicating a non-singlet state \cite%
{LB,YO,Th1,PAL1,PAL2,ST,ST1,EKH,HSZ,SCGO-2,SM}.

Figure 2(a) shows the typical temperature evolution of field swept $^{71}$Ga
NMR spectra of SGCO at a frequency $\nu $ = 69.5 MHz. With decreasing $T$,
although the $^{71}$Ga NMR spectra broaden, NMR shift $^{71}K$ for the main
line shows a broad maximum around 70 K, which is a characteristic feature of
low dimensional AFM spin systems due to short range spin correlations. Below
the broad maximum, $^{71}K$ decreases and levels off at low $T$ and then
remains nearly constant down to 50 mK. The frustration parameter ($f$\ ) is
considered to be a measure of the depth of the spin liquid regime and is
defined as $f$\textit{\ = }$\mid \theta _{CW}\mid $/$T_{N}$. In the present
case we did not observe magnetic ordering down to 50 mK, so $f$\ \textit{\ }$%
\geq \mid \theta _{CW}\mid $/50 mK $\sim $ 900 \cite{LB,AP1}. This suggests
the presence of strong magnetic frustration inspite of the large site
inversion. The frustration between Cu$^{2+}$ spins residing in the 2D
triangular bi-planes of SGCO might offer a route for the persistent spin
dynamics of Cu$^{2+}$ spins down to 50 mK and these fluctuating spins
preclude LRO. In addtion to the main $^{71}$Ga NMR line, we have observed a
weak line (labeled as Ga(II) in Fig. 2(a)) whose NMR\ shift $K_{\mathrm{II}}$
shows a CW behavior as shown in Fig. 2(b). Since the estimated signal
internsity of the Ga(II) line is 19 \% of the total $^{71}$Ga NMR\
intensity, which is in good agreement with an expected signal intensity of
which 20 \% Ga ions touching one Cu ion in the nearest neighbor of the
single layer, the Ga(II) signal can be ascribed to Ga ions in single layers.
The main Ga(I) line (Fig. 2(a)) is attributed to Ga ions in the triangular
bi-plane. We were not able to detect Ga(II) line and hence $K_{\mathrm{II}}$
at temperatures below 100 K due to inhomogeneous broadening of the spectra
perhaps because of Cu$^{2+}$ spins in the single triangular planes.

The NMR shift consists of $T$ dependent spin shift $K_{\mathrm{spin}}(T)$
and $T$ independent orbital (chemical) shift $K_{\mathrm{chem}}$; $K(T)$ = $%
K_{\mathrm{spin}}(T)$ + $K_{\mathrm{chem}},$ where $K_{\mathrm{spin}}(T)$ is
proportional to the spin part of magnetic susceptibility ${\chi _{\mathrm{%
spin}}(T)}$ via hyperfine coupling constant $A_{\mathrm{hf}}$, $K_{\mathrm{%
spin}}(T)$ = $A_{\mathrm{hf}}\chi _{\mathrm{spin}}(T)/N_{\mathrm{A}}$. Here $%
{N_{\mathrm{A}}}$ is Avogadro's number. The hyperfine coupling constant is
estimated to be $A_{\mathrm{hf}}$ = --3.8 $\pm $ 0.2 kOe/$\mu _{\mathrm{B}}$
for \ the main Ga(I) line from the slope of the so-called $K$-$\chi $ plots
using $\chi _{\mathrm{sub}}$ data at \textit{T} $\geq $ 150 K. $K_{\mathrm{%
chem}}$ \ values are estimated to be 0.049 $\%$ for the main Ga(I) line. The 
\textit{T}-dependence of the intrinsic magnetic susceptibility $\chi _{%
\mathrm{int}}$ obtained from $K_{\mathrm{spin}}$ data for the main line is
shown by solid spheres in Fig. 1(a). The $\chi _{\mathrm{int}}$ shows a
broad maximum around $\sim $ 70 K and decreases at low temperatures, but
does not approach zero. The nonzero value of $\chi _{\mathrm{int}}$ at low $%
T $ ($\sim $ 40 $\%$ of the maximum value) is strong evidence of the absence
of spin gap in SGCO. Similar behavior of $\chi _{\mathrm{int}}$ is reported
in the well known spin liquid material ZnCu$_{3}$(OH)$_{6}$Cl$_{2}$\cite%
{AO,TI}.

The $T$ dependence of $\chi _{\mathrm{int}}$ above $\sim $ 30 K is
reasonably reproduced by the high temperature series expansion (HTSE) of an$%
\ S$ = 1/2 triangular lattice Heisenberg model \cite{HTSE,KG} as shown in
Fig. 1(a) by the red line where the (4,7) Pad\'{e} approximant is adapted
with an effective exchange coupling between Cu$^{2+}$ spins with $J/k_{%
\mathrm{B}}$ = (35 $\pm $ 3) K (see Supplemental Material \cite{SM}). The
good fit indicates that, although more than 50\% of Cu$^{2+}$ ions in the
triangular bi-planes experienced site inversion, the intra-plane magnetic
interaction is still maintained. It should be noted that $\chi _{\mathrm{int}%
}$ does not coincide with $\chi _{\mathrm{sub}}$ at low $T$. This indicates
that the large enhancements of $\chi _{\mathrm{obs}}$ at low $T$ cannot be
explained only by the $\sim $ 15 \% Cu$^{2+}$ spins due to the antisite
effects. The exact origin for the difference between $\chi _{\mathrm{obs}}$
and $\chi _{int}$ is not clear at present but might be associated with the
site inversion between Cu and Ga sites in the system \cite{SCGO-2}. As shown
in Fig. 2(c), the full width at half maximum (FWHM = $\Delta H$) of the NMR
spectrum for the main line increases with decreasing $T$ and saturates below
2 K. The \ $T$-independent $\Delta H$\textit{\ }below 2 K is found to be
independent of the applied magnetic field indicating both $H$ and $T$%
-independent internal field at $^{71}$Ga sites. These results suggest that Cu%
$^{2+}$ spins fluctuate slowly \textit{i}.\textit{e}., at less than the NMR
frequency ($\sim $ 50 MHz) at low $T$. From the saturated $\Delta H$ value
at low \textit{T}, we estimated the Cu magnetic moments of magnitude 0.19 $%
\mu _{\mathrm{B}}$, which is quite small compared to the total magnetic
moment expected for $S$ = 1/2. The $^{45}$Sc NMR spectra, shift and $\Delta H
$ also exhibit a similar $T$-dependence with those of \ $^{71}$Ga NMR
results.

Figure 3 (a) depicts the $T$ dependence of spin-lattice relaxation rates$\
1/T_{1}$ of $^{71}$Ga, together with that of $^{45}$Sc. 1/$T_{1}$ is almost
independent of $T$ above 100 K and starts to decrease at low $T$ and then
levels off below $\sim $ 10 K down to 2 K. With further decreasing $T$, as
shown in Fig. 3(a), independent of probing nuclei, 1/$T_{1}$ decreases and
displays a power law behavior \textit{i.e}., 1/$T_{1}$ $\sim $ ${T}^{3.2}$
down to 100 mK. 1/$T_{1}$ is almost independent of magnetic field above 2 K,
but is suppressed strongly with magnetic fields at low $T$ as shown in the
Fig. 3(a).

A simple scenario for the decrease in 1/$T_{1}$ due to suppression of
magnetic fluctuations of isolated paramagnetic spins at high field and low $%
T $ cannot be attributed for the observed behavior. For the simple
paramagnetic spin fluctuations of isolated spins, 1/$T_{1}$ is known to be
proportional to the first derivative of the Brillouin function, $dB_{\mathrm{%
s}}(x)/dx$ ($x$=$g\mu _{B}S\mathit{H}/\mathit{k}_{B}\mathit{T}$) which gives
an exponential behavior of $1/T_{1}$ in $T$ following exp(-$g\mu _{B}\mathit{%
H}/\mathit{k}_{B}\mathit{T}$) function, in contrast to the power law
behavior in the observed $1/T_{1}$. As shown in Fig. 3(a), the exponent of
the power law in 1/$T_{1}$ is almost independent of magnetic fields implying
the intrinsic and robust nature of the ground state properties. It is worth
mentioning here that the power law dependence of spin-lattice relaxation
rate 1/$T_{1}\sim T^{\eta }$ has been discussed in the context of Dirac
Fermion model in interpreting QSL \cite{SYQ,PC,PAL1}. $1/T_{1}\sim T^{1.5}$
behavior in the $S$ = 1/2 triangular lattice $\kappa $-(BEDT-TTF)$_{2}$Cu$%
_{2}$(CN)$_{3}$ has been reconciled in the framework of $Z_{2}$ spin liquid
(SL) with quantum critical spin excitations \cite%
{FLP,SYQ,Th1,Th2,Th3,PAL2,TS}. Recently, another plausible theoretical
conjecture in interpreting the role of randomness in driving a gapless SL
state of $\kappa $-(BEDT-TTF)$_{2}$Cu$_{2}$(CN)$_{3}$ and EtMe$_{3}$%
Sb[Pd(dmit)$_{2}$]$_{2}$ is proposed \cite{KW,TF}. However, a general
consensus in interpreting the $T$ dependence of $1/T_{1}$ in the SL
materials is still lacking and little progress has been made in evolving a
more generic and comprehensive framework. This is due to the unavailability
of many model SL materials and experimental challenges in interpreting the
implications of various subtle theoretical scenarios \cite{LB,FLP,TF}.
Furthermore, one would expect a \textit{T}- independent behavior of 1/$%
T_{1}T $ in the case of a spin liquid with a spinon Fermi surface and 1/$%
T_{1}T$ should drop exponentially in the case of gapped SL \cite%
{LB,CL,Th1,Th2,Th3,TS,TD}. Our results are not in accord with the above
cited two scenarios but could be associated with the interpretation of not a
fully gapless SL where at least some part of the \textit{q}-space is gapped 
\cite{RK,KW}. In view of the power law behavior of magnetic specific heat
and 1/\textit{T}$_{1}$, a detailed theoretical investigation call for in
interepreting these results in the context of emergent excitations in the
gapless QSL, which is beyond the scope of the present study, but renders a
direction for further explorations \cite{ZY,Th1,PAL1,PAL2,ST,ST1,EKH}.

Finally, it is important to point out that our $T_{1}$ data indicate a
slowing down of Cu$^{2+}$ spin fluctuations at low temperature. 1/$T_{1}$ is
generally expressed by the Fourier transform of the time correlation
function of the transverse component $\delta $$h_{\pm }$ of the fluctuating
local field at nuclear sites with respect to the nuclear Larmor frequency $%
\omega _{\mathrm{N}}$ as \cite{BPP,Abragam} $\frac{1}{T_{1}}=\frac{\gamma _{%
\mathrm{N}}^{2}}{2}\int_{-\infty }^{+\infty }\langle \ h_{\pm }(t)h_{\pm
}(0)\ \rangle \mathrm{e}^{i\omega _{\mathrm{N}}t}dt$ , where $\gamma _{%
\mathrm{N}}$ is the gyromagnetic ratio of the nuclear spin. When the time
correlation function is assumed to decay as e$^{-\Gamma t}$, one can write $%
\frac{1}{T_{1}T\chi }=A\frac{\Gamma }{\Gamma ^{2}+\omega _{\mathrm{N}}^{2}}$
(eq.1) where $A$ is a parameter related to the hyperfine field at nuclear
sites and $\chi \ $is the magnetic susceptibility. In our case, $\Gamma $
would correspond to the inverse of the correlation time of the fluctuating
hyperfine fields at the Ga or Sc sites, due to the Cu$^{2+}$ spins. When $%
\Gamma $ is much higher than $\omega _{\mathrm{N}}$, one finds that the $%
\frac{1}{T_{1}T\chi }$ is proportional to 1/$\Gamma $. On the other hand, if 
$\Gamma $ $\ll $ $\omega _{\mathrm{N}}$, $\frac{1}{T_{1}T\chi }$ should
depend on the magnetic field. When $\Gamma $ = $\omega _{\mathrm{N}}$, $%
\frac{1}{T_{1}T\chi }$ reaches a maximum value. Thus, the slowing down of
the fluctuation frequency $\Gamma $ of Cu$^{2+}$ spins yields a peak in $%
\frac{1}{T_{1}T\chi }$. Figure 3(b) represents the temperature dependence of 
$\frac{1}{T_{1}T\chi }$, where the $\chi $ values are used for corresponding 
$K_{\mathrm{spin}}$ for each nucleus. When $\Gamma $ is independent of $T$,
1/$T_{1}TK_{\mathrm{spin}}$ should be constant, which is indeed observed
above 50 K. This indicates 1/$T_{1}$ above 50 K is explained by the
paramagnetic fluctuations of the Cu$^{2+}$ spins, whereby the Cu spins
fluctuate almost independently. Below 50 K, the 1/$T_{1}TK_{\mathrm{spin}}$
starts to increase and shows $H$ dependent peaks at low \textit{T} below 2
K. This can be explained by the slowing down in fluctuation frequency of
spins at low \textit{T}. These results indicate that the peak observed in 1/$%
T_{1}TK_{\mathrm{spin}}$ originates from the slowing down (but not critical)
of fluctuation frequency of Cu$^{2+}$ spins, whereby the fluctuation
frequency below the peak temperature is less than the NMR frequency range ($%
\sim $10 - 100 MHz). To derive the $T$ dependence of the fluctuation
frequency of Cu$^{2+}$ spins in a wide temperature range, we extract the $T$%
-dependence of $\Gamma $ from the $T$-dependence of 1/$T_{1}TK_{\mathrm{spin}%
}$, assuming eq. (1) is valid for all temperature regimes. The estimated $T$
dependences of $\Gamma $ for the different magnetic fields are shown in Fig.
4, together with the data estimated from $^{45}$Sc-$T_{1}$. $\Gamma $ shows $%
T^{2.2}$ behavior at low \textit{T} and is almost a constant with $\Gamma $ $%
\sim $ 3$\times $10$^{9}$ Hz at high $T$. At low \textit{T} below $\sim $ 1
K, Cu$^{2+}$ spins fluctuate with low frequency. Such a slow spin dynamics
is consistent with the observed broadening of the NMR spectra below $\sim $2
K. The absence of critical slowing down and no loss of NMR signal intensity
rule out the possibility of spin glass phase down to 50 mK in SGCO. This is
further substantiated by the absence of critical divergence of 1/\textit{T}$%
_{1}$ or cusp structure in 1/\textit{T}$_{1}$ generally observed in spin
frozen states\cite{TF}.

In summary, the intrinsic spin susceptibility ($\chi _{\mathrm{int}})$
obtained from NMR does not vanish and remains finite at \textit{T }= 50 mK,
reflecting a non-singlet ground state in Sc$_{2}$Ga$_{2}$CuO$_{7}$. The 
\textit{T}-dependence of $\chi _{\mathrm{int}}$ is well reproduced by the
HTSE of the \textit{S} = 1/2 Heisenberg model, indicating that the 2D
magnetic interactions between Cu$^{2+}$ spins in the bi-plane are still
maintained although more than 50\% spins involved in the unavoidable
intersite inversion. Quantum fluctuations enhanced by strong frustration
between Cu$^{2+}$ spins in the 2D triangular bi-plane inhibit the LRO down
to 50 mK despite an AFM exchange interaction $J/k_{\mathrm{B}}$ $\approx $
35 K. The spin-lattice relaxation rate exhibits a slowing down of \ Cu$^{2+}$
spin fluctuations and short range spin correlations at low $T$. The power
law behavior of \textit{C}$_{m}$ and 1/$T_{1}$ with decreasing temperature
down to 100 mK infer gapless excitations consistent with $\chi _{\mathrm{int}%
}$ \ and suggest a quantum spin liquid state. The effect of site dilution,
defect, and disorder in frustrated quantum magnets have been discussed in
the context of novel magnetism such as spin liquids recently \cite%
{BF,TF,FM,KW}. \ The absence of spin freezing and no spin gap down to 50 mK
in SGCO suggest that the low energy excitations might be mediated by
deconfined spinons, which is generic to a gapless QSL state in frustrated
quantum magnets. This point towards the predominant nature of deconfined
spinons in QSL state in case of the randomness induced by disorder due to
site inversion and frustration in realizing electron localization \cite%
{LB,TS1}. In this context SGCO offers a fertile ground for exploring the
effect of dilution or disorder, and the role of control parameters in tuning
emergent states in frustrated quantum magnets.

We acknowledge insightful discussions with P. Mendels, B. Koteswararao, and
B. Roy. PK acknowledges support from the European Commission through Marie
Curie International Incoming Fellowship (PIIF-GA-2013-627322). The research
was supported by the U.S. Department of Energy, Office of Basic Energy
Sciences, Division of Materials Sciences and Engineering. Ames Laboratory is
operated for the U.S. Department of Energy by Iowa State University under
Contract No. DE-AC02-07CH11358.

\qquad $\dagger $Present address: \ Laboratoire de Physique des Solides,
Universit\'{e} Paris-Sud 11, UMR CNRS 8502, 91405 Orsay, France.

*pkhuntia@gmail.com

\bigskip 

\bigskip \textbf{Figure Captions: }

\textbf{Fig. 1} (Color online) (a) Temperature dependence of the observed
magnetic susceptibility $\chi _{\mathrm{obs}}$ (solid line) at 7 T and the
subtracted magnetic susceptibility $\chi _{\mathrm{sub}}$ (dotted line)
after subtraction of 15 \% Cu spin contributions due to the site inversion
as discussed in the text. The solid spheres depict the intrinsic magnetic
susceptibility $\chi _{\mathrm{int}}$ estimated from $^{71}$Ga-NMR shift.
The red solid line is a fit as discussed in the text. (b) The inset shows
the \textit{T}-dependence of magnetic specific heat (\textit{C}$_{m}$) in
different magnetic fields and the solid line depicts the power law ($\sim $%
\textit{T}$^{1.9}$) behavior.

\textbf{Fig. 2} (Color online) (a) Temperature evolution of field swept $%
^{71}$Ga NMR spectra at 69.5 MHz. The vertical broken line corresponds to
zero-shift ($^{71}K$ = 0) position. (b) $T$ dependence of both $^{71}K$ for
main and Ga(II) lines. (c) $T$ dependence of NMR line width ($\Delta H$) at
69.5 MHz and 24.25 MHz.

\textbf{Fig. 3} (a) (Color online) Temperature dependence of $^{71}$Ga and $%
^{45}$Sc\ 1/$T_{1}$ at different frequencies. The solid line represents $%
T^{3.2}$ behavior (b) $T$ dependence of 1/$T_{1}TK_{\mathrm{spin}}$ (1/$%
T_{1} $ divided by temperature and respective spin susceptibilities $|$$%
^{45}K|$ and $|$$^{71}K|$).

\textbf{Fig. 4} (Color online) Temperature dependence of $\Gamma $\
estimated from $^{71}$Ga and $^{45}$Sc 1/$T_{1}$ as explained in the text.
The solid line is the $T^{2.2}$ behavior.

\subsection{\textbf{Methods}}

Polycrystalline Sc$_{2}$CuGa$_{2}$O$_{7}$ samples were synthesized by a
method described elsewhere \cite{SCGO-1}. Phase purity was confirmed by
Rietveld refinement of x-ray diffraction (XRD), synchrotron and neutron
diffraction data \cite{SCGO-1}. The temperature dependence of \textit{dc}
magnetic susceptibility $\chi _{\mathrm{obs}}$(=$M$($T$)/$H$) was measured
at 7 T in the temperature range 1.8 $\leq $ \textit{T} $\leq $ 400 K using a
Quantum Design, Physical Property Measurement System (PPMS). The absence of
hysteresis and spin-freezing were confirmed from the magnetization
measurements following the zero field and field cooled protocol\ in the
sample studied in this work. The low temperature specific heat measurment at
various applied magnetic fields was performed using the $^{3}$He option of
QD, PPMS following thermal relaxation method. The magnetic specific heat was
extracted from the measured specific heat by subtracting the lattice
specific heat and contribution from nuclear Schttoky \cite{SCGO-1}.The
exchange interaction between the nearest neighbor \ Cu$^{2+}$\ spins is
described by the Hamiltonian, $H$ = $J\sum_{<i,j>}$\ $S_{i}$.$S_{j}.$ In
order to estimate the exchange coupling ($J$) between the nearest neighbor Cu%
$^{2+}$ spins in the triangular biplane, we have fitted the intrinsic
magnetic susceptibility data following the high temperature series expansion
for $S$ = 1/2 system following the (4,7) Pad\'{e} approximant appropriate
for triangular lattice antiferromagnet

$\chi (T)=\frac{N_{A}g^{2}\mu _{B}^{2}}{k_{B}T}\sum_{n=0}^{11}\frac{a_{n}}{%
n!(4n+1)}(\frac{J}{k_{B}T})^{^{n}}$

where $a_{n}$ are series coefficients and the values of which can be found
in Ref.\cite{HTSE1,KG}. We obtained an exchange interaction of \ $J/k_{B}$ =
(35 $\pm $\ 3) K between Cu$^{2+}$ spins from the fit of intrinsic magnetic
susceptibility of Sc$_{2}$CuGa$_{2}$O$_{7}$.

Nuclear magnetic resonance (NMR) measurements down to 50 mK at various
frequencies were carried out on $^{71}$Ga ($I$ = 3/2, $\gamma /2\pi $ =
12.9847 MHz/T) and $^{45}$Sc ($I$ = 7/2, $\gamma /2\pi $ = 10.343 MHz/T) by
using a homemade phase-coherent spin-echo pulse spectrometer. The low
temperature NMR\ measurements are performed with a Oxford Kelvinox dilution
refrigerator installed at Ames Laboratory. NMR spectra were obtained by
sweeping the magnetic field $H$ at a fixed frequency. The temperature
dependence of NMR\ shift were obtained from the simulation of NMR\ spectra
taken at different temperatures. The $^{71}$Ga NMR\ spectrum at high
temperature were simulated with the superposition of two lines, one due to
Ga(I) and other due to Ga(II) because of defect Cu spins. This appears to be
due to antisite disorder between the Cu and the Ga atoms in the host
lattice. The ratio of NMR intensities for these two lines is estimated to be
0.81:0.19 at different temperatures, which implies 19 \% of Cu sits at the
Ga site and is consistent with those deduced from magnetization, specific
heat and neutron diffraction data. Site inversion between Cu and Zn is also
observed in the well known kagom\'{e} spin liquid material herbersmithite%
\cite{AO1,TI1}. The frustration parameter ($f$\ ), which is a measure of the
depth of spin liquid regime and is defined as $f$\textit{\ =}$\mid \theta
_{CW}\mid $/$T_{N}$, In the system presented here we didn't observe magnetic
ordering down to 50 mK, so $f$\ \textit{\ }$\geq \mid \theta _{CW}\mid $/50
mK $\sim $ 900\cite{LB1,AP1}. The large value of $f$\ indicates the presence
of strong magnetic frustration between Cu$^{2+}$ spins, which prevents LRO
down to 50 mK and leads to exotic magnetic properties in Sc$_{2}$CuGa$_{2}$O$%
_{7}$ discussed here.

Furthermore, the role of perturbations such as lattice disorder due to site
inversion, presence of defect spins and/or nonmagnetic substituents in
frustrated antiferromagnets in modifying local environments by inducing
magnetic moments in its immediate vicinity lead to interesting magnetic
properties\cite{AO1,TI1,Nmat,AVM,MAV,FM1,HA,RRP}.

In order to investigate dynamics of the Cu spins and the ground state
properties, we have performed spin-lattice relaxation rate (1/$T_{1}$)
measurements in the wide temperature range 0.1 $\leq T\leq $ 250 K for the
central line. The recoveries of the longitudinal magnetization for both
nuclei display stretched exponential behavior suggesting distributions of $%
T_{1}$ values. The $1/T_{1}$ at each $T$ is determined by fitting the
nuclear magnetization $M$($t$) using the stretched double-exponential
function 1-$M$($t$)/$M({\infty })$ = 0.1e$^{-(t/T_{\mathrm{1}})^{\mathrm{%
\beta }}}$+ 0.9e$^{-(6t/T_{\mathrm{1}})^{\mathrm{\beta }}}$ for the central
line of the spectrum of the $^{71}$Ga ($I$ = 3/2) nucleus\cite{AN1}. The
recovery of \ $M$($t$) at all temperatures could be fitted with $\beta $ $%
\approx $ 0.5 ($\beta $ is found to be nearly independent of temperature).
Here $M$($t$) and $M({\infty })$ are the nuclear magnetization at time $t$
after saturation and the equilibrium nuclear magnetization at time $t$ $%
\rightarrow $ $\infty $, respectively. Similarly, for $^{45}$Sc ($I$ = 7/2
), spin-lattice relaxation rate was obtained by fitting $M$($t$) using
stretched exponential function 1-$M$($t$)/$M({\infty })$ = 0.0119e$^{-(t/T_{%
\mathrm{1}})^{\mathrm{\beta }}}$+ 0.0682e$^{-(6t/T_{\mathrm{1}})^{\mathrm{%
\beta }}}$+ 0.206e$^{-(15t/T_{\mathrm{1}})^{\mathrm{\beta }}}$+ 0.7139e$%
^{-(28t/T_{\mathrm{1}})^{\mathrm{\beta }}}$ ($\beta $ $\approx $ 0.5) valid
for $I$ = 7/2\cite{AN1}.

We believe that the present work should open new avenues in frustrated
magnetism and will stimulate further theoretical and experimental
investigations exploring the nature of low lying excitations and the role of
perturbations on the ground state of triangular lattice antiferromagnets.

*pkhuntia@gmail.com

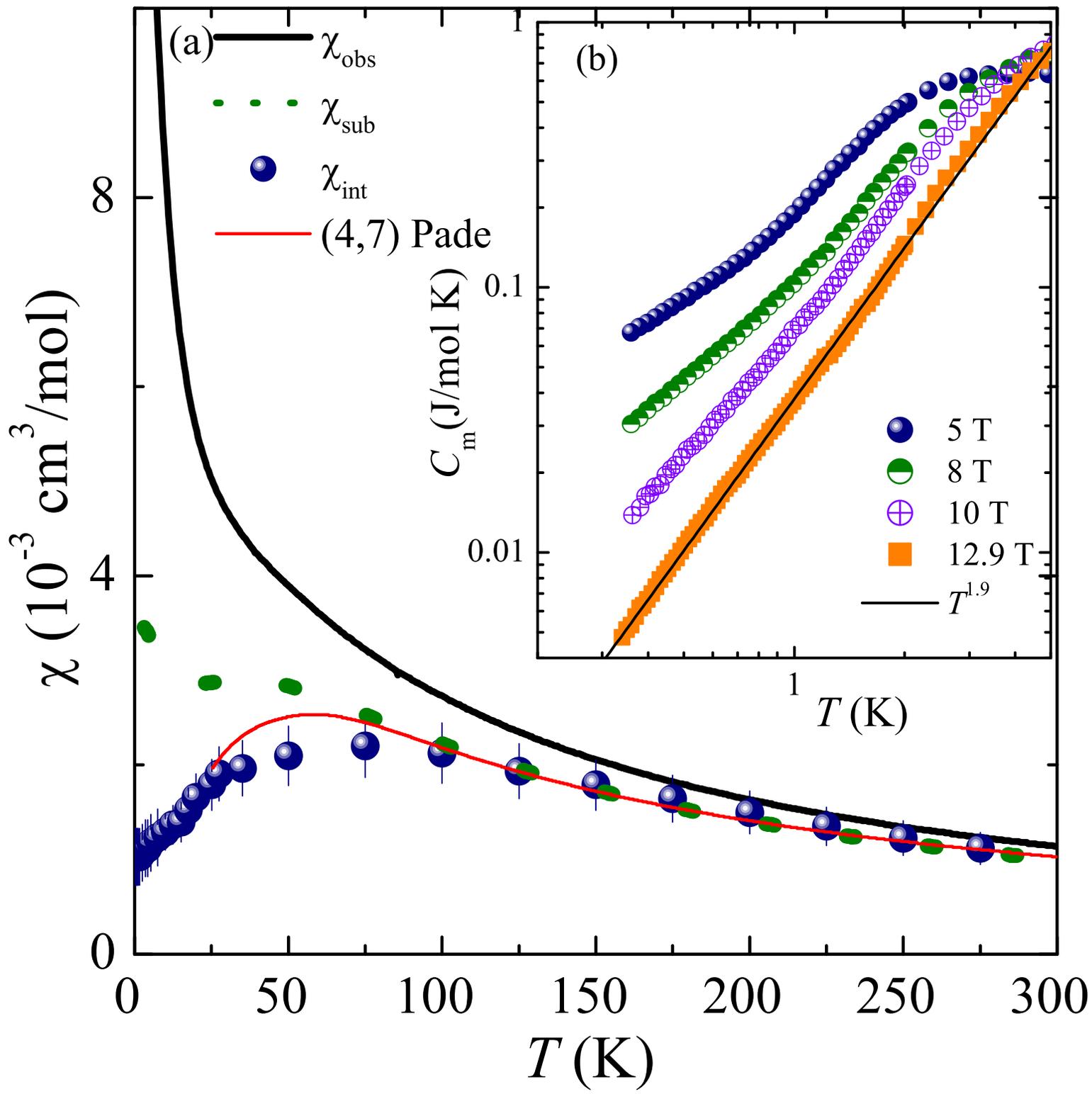
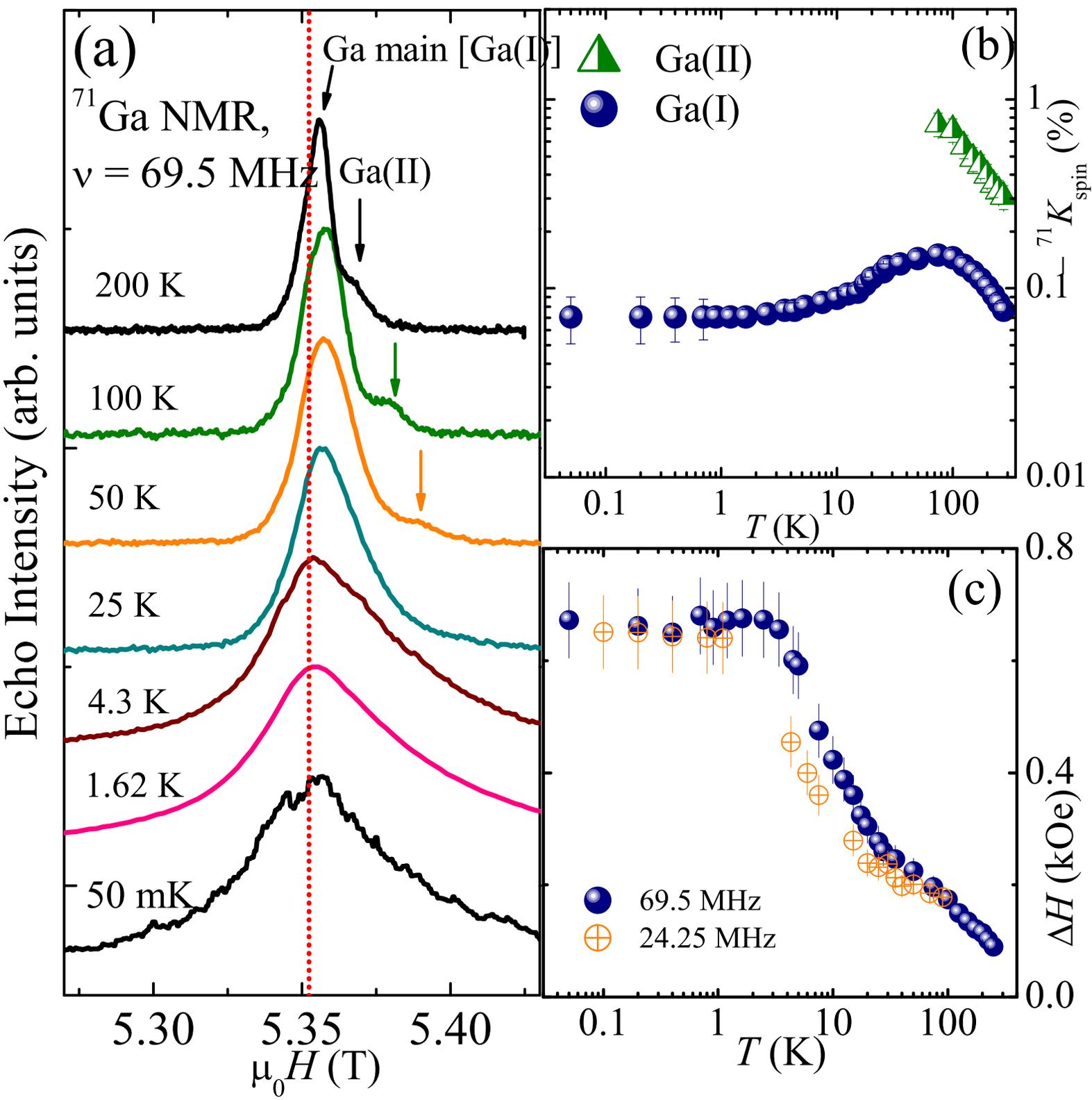
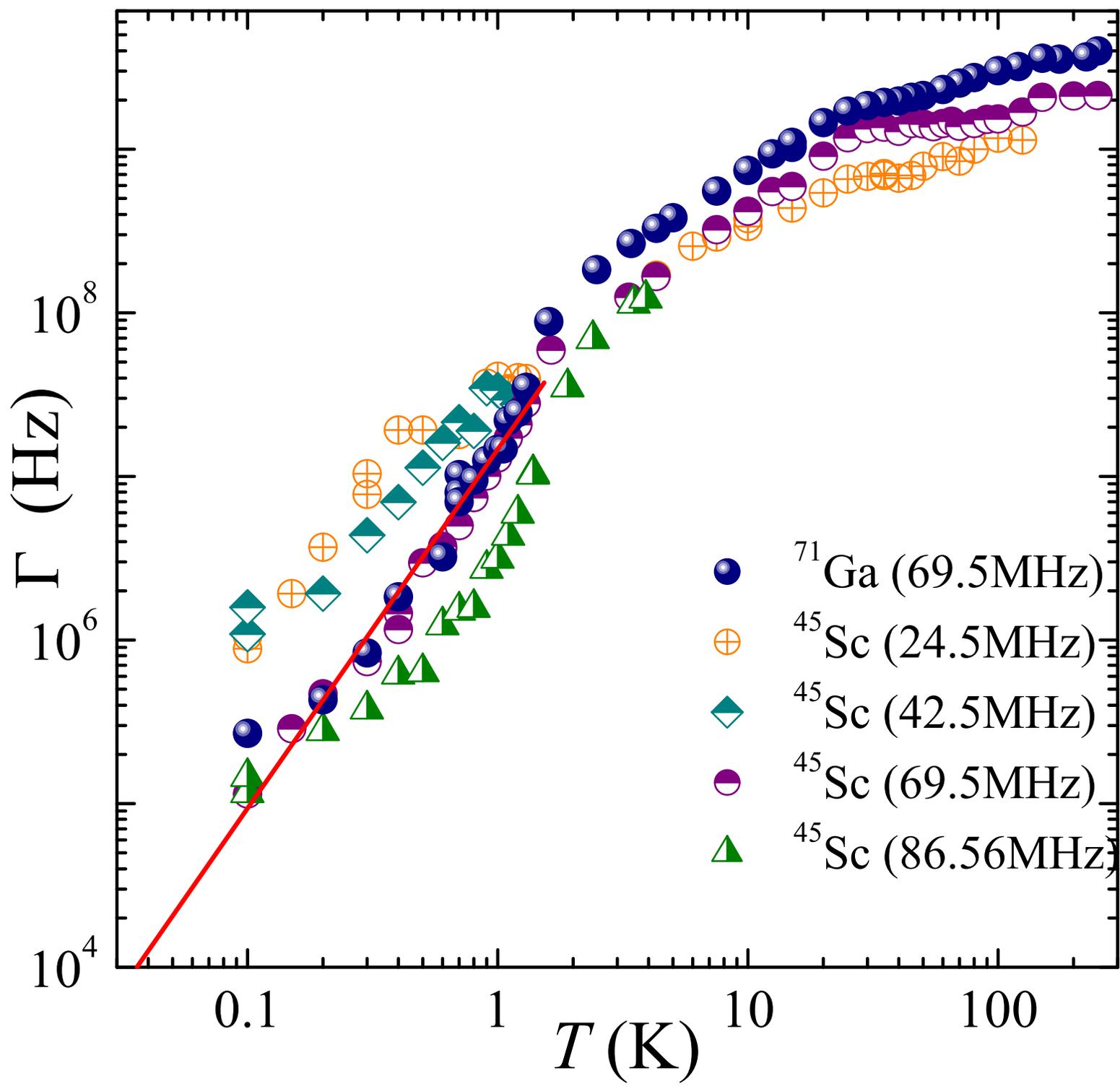
\end{document}